\documentclass[runningheads]{llncs}
\usepackage{graphicx}
\usepackage{romannum}
\usepackage{booktabs}
\usepackage[sort]{cite}
\usepackage{subcaption}
\usepackage{amsmath,esint}
\begin{document}
\title{Deep Learning for Hindi Text Classification : \\A Comparison}
%
%
\author{Ramchandra Joshi$^1$ \and
Purvi Goel$^2$ \and
Raviraj Joshi$^2$}
\authorrunning{R. Joshi et al.}
%
\institute{$^1$Department of Computer Engineering, Pune Institute of Computer Technology\\
$^2$Department of Computer Science and Engineering, Indian Institute of Technology Madras\\
\email{\{rbjoshi1309, goyalpoorvi, ravirajoshi\}@gmail.com}}
\maketitle              
\begin{abstract}
Natural Language Processing (NLP) and especially natural language text analysis have seen great advances in recent times. Usage of deep learning in text processing has revolutionized the techniques for text processing and achieved remarkable results. Different deep learning architectures like CNN, LSTM, and very recent Transformer have been used to achieve state of the art results variety on NLP tasks.
In this work, we survey a host of deep learning architectures for text classification tasks. The work is specifically concerned with the classification of Hindi text. The research in the classification of morphologically rich and low resource Hindi language written in Devanagari script has been limited due to the absence of large labeled corpus. In this work, we used translated versions of English data-sets to evaluate models based on CNN, LSTM and Attention. Multilingual pre-trained sentence embeddings based on BERT and LASER are also compared to evaluate their effectiveness for the Hindi language. The paper also serves as a tutorial for popular text classification techniques.

\end{abstract}
\section{Introduction}
Natural language processing represents computational techniques used for processing human language. The language can either be represented in terms of text or speech. NLP in the context of deep learning has become very popular because of its ability to handle text which is far from being grammatically correct. Ability to learn from the data have made the machine learning system powerful enough to process any type of unstructured text. Machine learning approaches have been used to achieve state of the art results on NLP tasks like text classification, machine translation, question answering, text summarization, text ranking, relation classification, and others.
\\
The focus of our work is text classification of Hindi language. Text classification is the most widely used NLP task. It finds application in sentiment analysis, spam detection, email classification, and document classification to name a few. It is an integral component of conversational systems for intent detection. There have been very few text classification works in literature focusing on the resource-constrained Hindi language. While the most important reason for this is unavailability of large training data; another reason is generalizability of deep learning architectures to different languages. However, Hindi is morphologically rich and relatively free word order language so we investigate the performance of different models on Hindi text classification task. Moreover, there has been a substantial rise in Hindi language digital content in recent years. Service providers, e-commerce industries are now targeting local languages to improve their visibility. Increase in the robustness of translation and transliteration systems have also contributed to the rise of NLP systems for Hindi text. This work will help in the selection of right models and provide a suitable benchmark for further research in Hindi text classification tasks.
\\
In order to create Hindi dataset in Devanagari script, standard datasets like TREC, SST were translated using Google translate. This is indeed great times for translation systems. Self attention-based models like transformer have resulted in best in class results for translation tasks. We believe this is the right time to evaluate the translated data sets. Even the multi-lingual datasets like XNLI \cite{conneau2018xnli} used for evaluation of natural language inference tasks is based on the translation.
\\
Current text classification algorithms are mainly based on CNNs and RNNs. They work at the sentence level, paragraph level or document level. In this work, we consider sentence-level classification tasks. Each sentence split into a sequence of word tokens and passed to classification algorithms. Instead of passing the raw character tokens, each word is mapped to a numerical vector. The sequence of vectors is processed by classification algorithms. A very common approach is to learn these distributed vectorial representations using unsupervised learning techniques like word2vec \cite{pennington2014glove}. The similarity of the word vectors are correlated to the semantic similarity between actual words. This gives some useful semantic properties to low dimensional word vectors. Usage of pre-trained vectors have shown to give superior results and are thus the de-facto method to represent word tokens in all NLP models.  In this work, we use FastText word vectors, pre-trained on Hindi corpus. The embedding matrix is used as an input to deep learning models. Naive bag of words approach is to average the word embeddings and then use a linear classifier or feed-forward neural network to classify the resulting sentence embedding. A more sophisticated approach is to pass the sequence of word vectors through LSTM and use final hidden state representation for classification. CNNs are also pretty popular for sentence classification tasks where a fixed length padded word vector sequence is passed through the CNNs. We explore different variations of LSTM, CNN, and Attention-based neural networks for comparison.
\\
Learning universal sentence representations is another area of active research. These sentence representations are used for classification tasks. General idea is to use a large amount of labeled or unlabelled corpus to learn sentence representations in a supervised or unsupervised setting. This is similar to learning word vectors externally and using them in the target task. These approaches represent transfer learning in the context of NLP. Models like Skip-Thought Vectors, Universal Sentence Encoder by Google, InferSent, and BERT have to be used to learn sentence embeddings. Using pre-trained sentence embeddings lowers the training time and is more robust on small target data sets. In this work, we also evaluate pre-trained multi-lingual sentence embedding obtained using BERT and LASER to draw a better comparison. 
\\
Main contributions of this paper are:
\begin{itemize}
    \item Compare variations of CNN and LSTM models for Hindi text classification.
    \item Effectiveness of Hindi Fast-Text word embedding is evaluated. 
    \item Effectiveness of multi-lingual pre-trained sentence embedding based on BERT and LASER is evaluated on Hindi corpus.
\end{itemize}
\section{Related Work}

There has been limited literature on Hindi text classification. Arora Piyush in his early work \cite{arora2013sentiment} used traditional n-gram and weighted n-grams method for sentiment analysis of Hindi text. Tummalapalli \textit{et al.} \cite{tummalapalli2018towards} used deep learning techniques- basic CNN, LSTM and multi-Input CNN for evaluating the classification accuracy of Hindi and Telugu texts. Their main focus was capturing morphological variations in Hindi language using word-level and character-level features. CNN based models performed better as compared to LSTM and SVM using n-gram features. The datasets used were created using translation. In this work, we are concerned with the performance of different model architectures and word vectors so we do not consider character level or subword level features.
\\
In general, there has been a lot of research on text classification and sentiment analysis employing supervised and sem-supervised techniques. Kim \textit{et al.} \cite{kim2014convolutional} proposed CNN based architecture for classification of English sentences. A simple bag of words model based on averaging of fast text word vectors was proposed in \cite{joulin2016bag}. They proposed a simple fast baseline for sentence classification tasks. Usage of RNNs for text classification was introduced in \cite{lai2015recurrent} and Bi-LSTM was augmented with simple attention in \cite{zhou2016attention}. Classification results of these models on Hindi text are reported in this work.
\\
Sentence embeddings evaluated in this work include multi-lingual LASER embeddings \cite{artetxe2018massively} and multi-lingual BERT based embeddings \cite{devlin2018bert}. LASER uses Bi-LSTM encoder to generate embeddings whereas BERT is based on Transformer architecture. LASER takes a neural machine translation approach for learning sentence representations. It builds a sequence to sequence model using Bi-LSTM encoder-decoder architecture. The encoder Bi-LSTM is used to generate sentence representations. BERT, on the other hand, uses bi-directional transformer encoder for learning word and sentence representations. It uses masked language model as the pre-training objective to mitigate the problem of unidirectional training in simple language model next word prediction task.     
\section{Datasets}
\begin{itemize}
    \item TREC question dataset which involves classifying a question sentence into six types. The dataset has predefined train-test split. It has 5452 training samples and 500 testing samples. 10 \% of the training data was randomly held out for validation. 
    \item Stanford Sentiment Treebank datasets SST-1 and SST-2. SST-1 contains one sentence movie reviews which are rated in the scale of 1-5 going from positive to negative. The dataset has predefined train-test-dev split. It has 8544 training samples, 2210 testing samples, and 1101 validation samples. SST-2 is a binary version of SST-1 where there are only two labels positive and negative. It has 6920 training samples, 1821 testing samples, and 872 validation samples.
\end{itemize}
Original English versions of this dataset are translated to Hindi using Google Translate. A language model was trained using Hindi wiki corpus and used to filter out noisy sentences. We assume no out of vocabulary words as fast text model generates word embeddings for unknown words as well. A common vocabulary of 31k words is created and fast-text vectors are used to initialize the embedding matrix.
\section{Model Architectures}
The data samples comprise of a sequence of words so different sequence processing models are explored in this work. While the most natural sequence processing model is LSTM, other models are equally applicable as the sequence length is short. 
\begin{itemize}
    \item \textbf{BOW}: The bag of words model does not consider the sequence of words. The word vectors of input sentence are averaged to get a sentence embedding of size 300. This is followed by a dense layer of size equal to the number of output classes. Softmax output is given to cross-entropy loss function and Adam is used as an optimizer.
    \item \textbf{BOW + Attention}: In this model, instead of simply averaging, a weighted average of word vectors is taken to generate sentence embedding. The size of sentence embedding is 300 and is followed by a dense layer similar to BOW model. The weights for the individual time step is learned by passing the corresponding word vector through a linear layer of size $300 \times 1$. Softmax over these computed weights gives the probabilistic attention scores. This attention approach is described in \cite{zhou2016attention}.
    \item \textbf{CNN}: The sequence of word embeddings are passed through three 1D convolutions of kernel sizes 2, 3, and 4. Each convolution uses a filter size of 128. The output of each of the 1-D convolution is max pooled over time and concatenated to get the sentence representation. The size of this sentence representation is 384 dimensions. There is a final dense layer of size equal to the number of output classes.
    \item \textbf{LSTM}: The word vectors are passed as input to two-layer stacked LSTM. The output of the final time step is given as an input to a dense layer for classification. LSTM cell size is 128 and the size of final time step output which is treated as sentence representation is 128. 
    \item \textbf{Bi-LSTM}: The sequence of word embedding is passed through two stacked bi-directional LSTM. The output is max pooled over time and followed by a dense layer of size equal to the number of output classes. LSTM cell size is 128 and the size of max-pooled output which is treated as sentence representation is 256.
    \item \textbf{CNN + Bi-LSTM}: The sequence of word embeddings are passed through a 1D convolution of kernel size 3 and filter size 256. The output is passed through a bi-directional LSTM. The output of Bi-LSTM is max pooled over time and followed by a final dense layer.
    \item \textbf{Bi-LSTM + Attention}: This is similar to Bi-LSTM model. The difference is that instead of max-pooling over the output of Bi-LSTM an attention mechanism is employed as described above.
    \item \textbf{LASER and BERT}: Single pre-trained model for learning multilingual sentence representations in the form of BERT and LASER was released by Google and Facebook respectively. BERT is a 12 layer transformer based model trained on multilingual data of 104 languages. LASER a 5 layer Bi-LSTM model pre-trained on multilingual data of 93 languages. Both of these models have Hindi as one of the training languages. The sentence embeddings extracted from these models are used without any fine-tuning or modifications. The pre-trained sentence embeddings are extracted from the corresponding models and subjected to a dense layer of 512 units. It is further connected to a dense layer of size equal to the number of output classes over which softmax is computed. BERT generated 768-dimensional embedding whereas the dimension of LASER embeddings were 1024.
\end{itemize}
\section{Results and Discussion}
Performance of different models based on CNN and LSTM were evaluated on translated versions of TREC, SST-1, and SST-2 datasets. Different versions of input word vectors were given to the models for comparison. Pre-trained fast text embeddings trained on Hindi corpus were compared against random initialization of word vectors. The random values were sampled from a continuous uniform distribution in a half-open interval [0.0, 1.0). Moreover, in one setting pre-trained fast-text embeddings were fine-tuned whereas in other settings they remained static. Keeping the word vector layer un-trainable allows better handling of words that were not seen during training as all the word vectors follow the same distribution. However, the domain of the corpus on which the word vectors were pre-trained may be different from the target domain. In such cases, fine-tuning the trained word vectors helps model adapt to the domain of the target corpus. So re-training the fast text vectors and keeping them static has its pros and cons. Table \ref{tab1} shows the results of the comparison. The three versions of word vectors are indicated as random for random initialization, fast text for trainable fast-text initialization, and fast text-static for un-trainable fast text initialization. Out of all the models vanilla CNN performs the best for all the datasets. CNNs have known to perform best for short texts and same is visible here as the datasets under consideration do not have long sentences. There is a small difference in the performance of different LSTM model. However, Bi-LSTM with max-pooling performed better than its attention version and unidirectional LSTM. Bag of words based on attention fared better than the simple bag of words model. Attention was particularly helpful with the usage of static fast text word vectors.  Stacked CNN-LSTM models were somewhere between LSTM and CNN based models. We did not see a huge drop in performance due to random initialization of word vectors. But the performance across different epochs was very stable with fast text initialization. Finally, as compared to generic sentence embeddings obtained from BERT and LASER, specific embeddings obtained from custom models performed better. LASER was able to reach close to the best performing model. This shows that LASER was able to capture important discriminative features of a sentence required for the task at hand whereas  BERT failed to capture the same. 
\begin{table}
\centering
\renewcommand{\arraystretch}{1.0}
\renewcommand{\tabcolsep}{2mm}
\caption{Classification accuracies of different models}\label{tab1}
\begin{tabular}{|l|l|l|l|l|}
\hline
Model / Dataset & & TREC & SST-1 & SST-2\\
\hline
BOW & fast text-static & 62.4  & 32.2  & 63.1  \\
 & fast text & 87.2  & 40.4  & 77.3  \\
 & random & 84.4  & 39.3  & 76.9  \\
BOW-Attn & fast text-static & 76.2  & 37.4  & 72.8  \\
 & fast text & 88.2  & 39.3  & 78.0  \\
 & random & 86.0  & 36.9  & 75.4  \\
LSTM & fast text-static & 86.6  & 40.2  & 75.5  \\
 & fast text & 87.8  & 40.8  & 78.1  \\
 & random & 86.8  & 40.7  & 76.8  \\
Bi-LSTM & fast text-static & 87.0  & 40.8  & 76.4  \\
 & fast text & 89.8  & 41.9  & 78.0  \\
 & random & 87.6  & 40.2  & 72.9  \\
Bi-LSTM-Attn & fast text-static & 85.0  & 39.0  & 76.4  \\
 & fast text & 88.6  & 40.1  & 78.6  \\
 & random & 86.0  & 39.5  & 76.0  \\
CNN & fast text-static & 91.2 & 41.2  & 78.2  \\
 & fast text & \textbf{92.8}  & \textbf{42.9}  & \textbf{79.4}  \\
 & random & 87.8  & 40.2  & 77.1  \\
CNN+Bi-LSTM & fast text-static & 89.6  & 40.4  & 78.3  \\
 & fast text & 90.5  & 41.0  & 77.4  \\
 & random & 87.6  & 38.2  & 72.3  \\
LASER &  & 89.0  & 41.4  & 75.9  \\
BERT &  & 77.6  & 35.6  & 68.5  \\
\hline
\end{tabular}
\end{table}

\section{Conclusion}
In this work, we compared different deep learning approaches for Hindi sentence classification. The word vectors were initialized using fast text word vectors trained on Hindi corpus and random word vectors. This work also serves the evaluation of fast text word embeddings for Hindi sentence classification task. CNN models perform better than LSTM based models on the datasets considered in this paper. Although we would expect BOW to perform the worst it has numbers comparable to LSTM and CNN. Therefore if we can trade off accuracy for speed BOW is useful. LSTMs do not do better than CNNs may be because the word order is relaxed in Hindi. Sentence representations captured by LASER multilingual model were rich as compared to BERT. However, overall custom trained models on specific datasets performed better than lightweight models directly utilizing sentence encodings. Although the real advantage of multi-lingual embeddings can be better evaluated on tasks involving text from multiple languages.

%
%
%
%
\bibliographystyle{splncs04}
\bibliography{main}
\end{document}